\documentclass{article}
\pdfoutput=1
\usepackage{spconf,amsmath,graphicx,multirow}

\title{Streaming on-device detection of device directed speech \\ from voice and touch-based invocation}

\name{Ognjen (Oggi) Rudovic*, Akanksha Bindal*, Vineet Garg*, Pramod Simha*, Pranay Dighe, Sachin Kajarekar\thanks{*These authors contributed equally to this work. We also thank 
A.H. Abdelaziz, E. Marchi, C. Dhir and C. Mannemala for their support during the course of this work.}}
\address{Apple}
\begin{document}
%
\maketitle
\begin{abstract}

When interacting with smart devices such as mobile-phones or wearables, the user typically invokes a virtual assistant (VA) by saying a keyword or by pressing a button on the device. However, in many cases, the VA can accidentally be invoked by the keyword-like speech or accidental button press, which may have implications on user experience and privacy. To this end, we propose an acoustic false-trigger-mitigation (FTM) approach for on-device device-directed speech detection that simultaneously handles the voice-trigger and touch-based invocation. To facilitate the model deployment on-device, we introduce a new streaming decision layer, derived using the notion of temporal convolutional networks (TCN)~\cite{bai2018empirical}, known for their computational efficiency. To the best of our knowledge, this is the first approach that can detect device-directed speech from more than one invocation type in a streaming fashion. We compare this approach with streaming alternatives based on vanilla Average layer, and canonical LSTMs, and show: (i) that all the models show only a small degradation in accuracy compared with the invocation-specific models, and (ii) that the newly introduced streaming TCN consistently performs better or comparable with the alternatives, while mitigating device-undirected speech faster in time, and with (relative) reduction in runtime peak-memory over the LSTM-based approach of $33\%$ vs. $7\%$, when compared to a non-streaming counterpart. 



\end{abstract}
\section{Introduction}
\label{sec:intro}

Smart devices, such as mobile phones, speakers, wearables, and other AI-powered appliances and spaces (e.g. the voice-controlled rooms), are becoming common in daily life. To facilitate interaction, usually virtual assistants (VAs) are at the core of those devices, seamlessly connecting the user with device functions~\cite{mallidi2018devicedirected}. This enables the user to use voice commands and/or other types of invocation (e.g. hand-claps or pressing a button on a mobile phone) to initiate and control the interaction with target devices. Therefore, to achieve natural, non-intrusive and privacy-centric interaction between the user and VA, it is important that user intent to communicate with VA is accurately recognized by the device. 



To mitigate unintended invocations, most VAs rely on FTM systems designed to detect if the user's speech was device-directed~\cite{mallidi2018devicedirected}. In case of voice-trigger (VT) invocation, this is typically referred to as VT detection (VTD), also known as keyword spotting~\cite{sainath2015convolutional}, wake-up/hot -word detection~\cite{wu2018monophone, sigtia2020progressive, guo2018time}. Most existing FTM algorithms for this type of invocation rely on accurate detection of a keyword: using either a single pass low-powered approach (i.e., looking only for the keyword) or two-pass algorithms (where the candidate audio trigger is segmented from the first pass and further analyzed, with additional audio content ({\it payload}) following the keyword~\cite{sigtia2020progressive}). Even though VT utterances begin with trigger-word's phone sequence (see Table~\ref{tab:example}), accurate detection of the trigger-word(s) is difficult, mainly due to possible confusing words and acoustically challenging environments~\cite{Sigtia_2020}. Conversely, touch-based (TB) invocations may also falsely trigger the device (e.g., by accidentally pressing the button on a smart-phone). In such cases, FTM is more challenging as the VA cannot rely on detection of a keyword and it needs to deduce from the user's speech, which can begin with any word, if it was device-directed or not~\cite{mallidi2018devicedirected}.

\begin{table}[]
\centering
\caption{Examples of device (un)directed utterances.}
\resizebox{0.65\textwidth}{!}{\begin{minipage}{\textwidth}
\begin{tabular}{|l|l|l|}
\hline
 {\it Invocation}  &Device-directed   &Device-undirected  \\ \hline
\multirow{2}{*}{Voice-trigger (VT)} &"{\bf hey Siri}, set alarm for 2:15PM"  & "... history repeats itself..." \\ 
                  &"{\bf hey Siri}, call Mom"  &"... hey Sharon ..."  \\ \hline
\multirow{2}{*}{Touch-based (TB)} &"Taxi cab service near me"  &"... I am not that hungry now ..." \\ 
                  &"NBA playoffs"  &"... they said they would help ..."  \\ \hline
\end{tabular}
\end{minipage}}
\label{tab:example}
\end{table}



To improve the user experience (e.g., by reducing the response latency, which depends on the speech processing time) and address privacy concerns, most recent speech processing tasks are on-device. This has been enabled by rapid progress in the underlying hardware architectures, and the streaming design of target algorithms. Existing on-device streaming solutions focus mainly on tasks such as speech recognition~\cite{cao21b_interspeech, tsunoo2019transformer, oh21_interspeech}, and speech synthesis~\cite{wu2021transformer,ellinas2020high}. In the context of FTMs, it has only recently been proposed for the VT invocation (e.g., see ~\cite{tcn4kws, wang2021wake, garg2021streaming, dighe2020knowledge, Tong2020StreamingRW}) to be on-device. 
The role of the streaming mechanism is not only to minimize the runtime memory/latency, but also provide mitigation decisions at fixed time intervals as audio arrives. While processing on the server can be done with abundant compute resources, on-device FTM needs to satisfy hardware constraints (the asset size and runtime memory/latency). This requires a careful design of the algorithms that achieve the best accuracy and on-device performance simultaneously. 

\begin{figure}[htb]
\centering
\includegraphics[width=8.5cm]{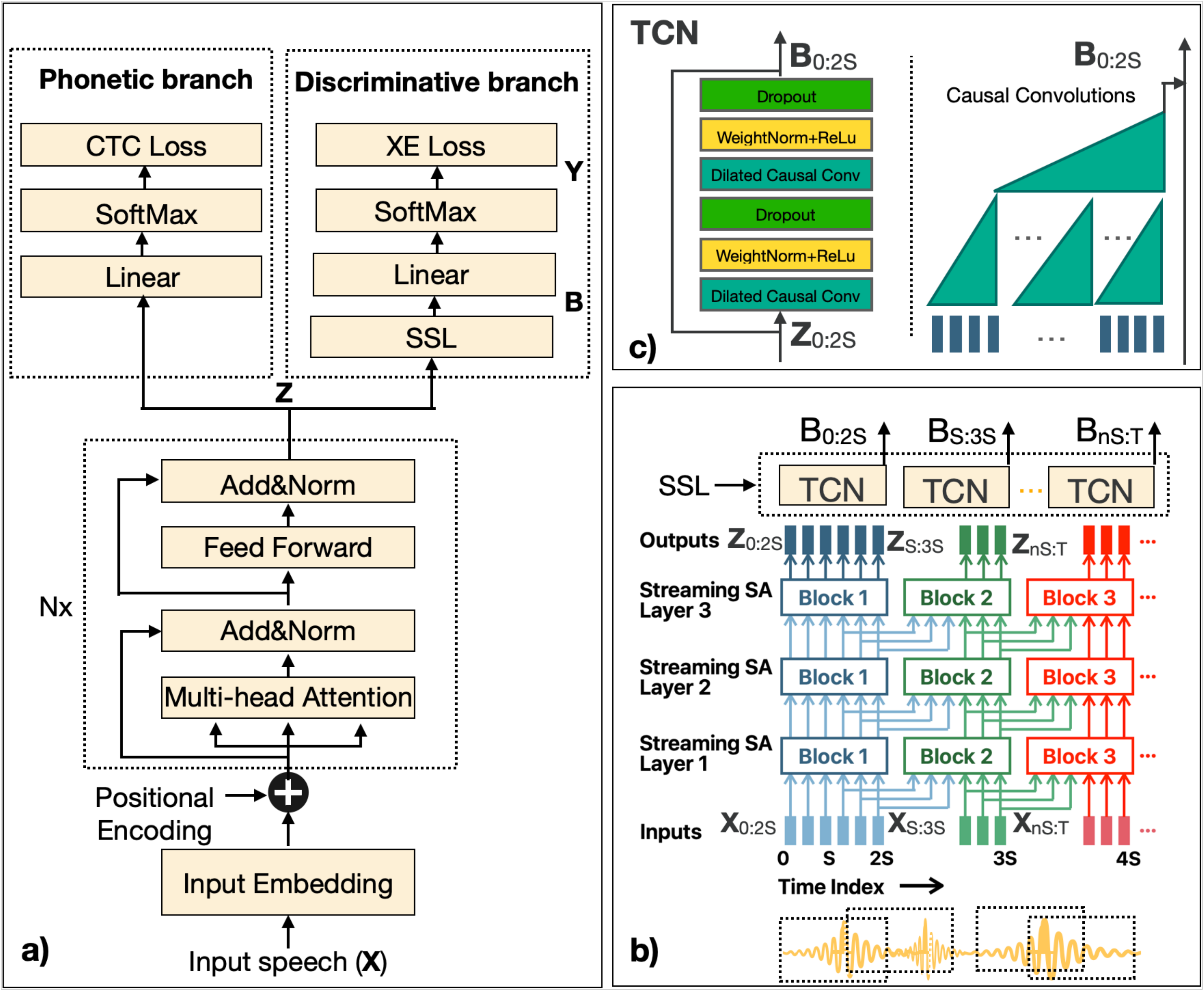}
\caption{{\bf System overview.} a) The adopted FTM architecture from~\cite{garg2021streaming}, with attention layers~\cite{vaswani2017attention}, acting as high-level feature extractors from input sound ($X$). The sequence summary layer (SSL) is used for the mitigation task. b) The proposed s-TCN combines the output of the {\it streaming} attention layers (Sec.~\ref{sec:stcn}) with the block-wise processing of the audio embeddings ($Z$) using the TCN residual unit (c). The block-wise scores ($B$) are averaged over time to produce the mitigation score for the target utterance ($Y$).}
\label{fig:system}
\end{figure}




\section{Baseline System Overview}
\label{sec:sa}
We adopt the architecture of a multi-task system depicted in Fig.\ref{fig:system} (a), and recently proposed for the VTD task~\cite{garg2021streaming, Sigtia_2020}. This architecture is derived from transformer networks~\cite{dai2019transformerxl}, where encoder is built from the multi-head attention layers~\cite{vaswani2017attention}, which serve here as the sequence-level audio feature extractors. For VTD, the multi-task network implements the keyword checker based on the decoded phoneme sequence in the phonetic branch. The discriminative branch performs FTM on audio segments that triggered the first pass. The network is optimized simultaneously on these two tasks. In this work, the  discriminative (FTM) branch is used for both the VT and TF invocations, while the phonetic branch serves as a regularizer during model training. The connectionist temporal classification (CTC) loss~\cite{NEURIPS2018_e44fea3b} is used in the phonetic branch~\cite{adya2020hybrid}. In the discriminative branch, the sequence summary layer (SSL) is implemented using a unidirectional LSTM with the frame-wise cross-entropy (XE) loss. To enable VTD in a streaming fashion, this architecture was modified by replacing the vanilla architecture with streaming attention (SA) layers through a masking mechanism (see Sec 2.2.1 in~\cite{garg2021streaming}). The authors in~\cite{garg2021streaming} showed that the streaming network matches the all-to-all (A2A), i.e. no streaming, model accuracy in the VTD task. In the following, we introduce s-TCN, an alternative streaming approach that combines the SA layers and TCN, to obtain FTM decisions over time. 
\section{Streaming TCN (s-TCN)}
\label{sec:stcn}
We explore an alternative summarization mechanism for our streaming approach (SSL). Recently reported empirical evaluations of TCN, comparing canonical recurrent models including LSTMs~\cite{bai2018empirical, tcn4kws, lea2016temporal}, have shown that a simple TCN is more effective across diverse sequence (language) modeling tasks in terms of compute parallelism, and training/inference memory, both of which are important for on-device deployment. The basic idea of TCNs is that they combine the concept of 1D fully convolutional networks (FCN)~\cite{kiranyaz20211d} and causal convolutions, where an output at time $t$ is convolved only with elements from time $t$ and earlier~\cite{bai2018empirical}. 

We illustrate s-TCN in Fig.~\ref{fig:system}(b). The incoming input audio (speech) is passed through the SA layers (in this example, $N=3$), where the processing is done in a block-wise manner (block size= $2S$), with an overlap of $S$ frames to allow for context propagation\footnote{This is implemented by applying a diagonal attention mask to the vanilla self-attention layers~\cite{garg2021streaming}.}. Note that the output embeddings ($Z$) are also computed in a causal manner, no leakage from future into the past is allowed. The outputs of each block are then fed into the TCN residual unit, the base architecture of which is depicted in Fig.~\ref{fig:system}(c). More formally, each TCN unit consumes the block-wise sequence of audio embeddings, and outputs the sequence of embeddings the same length as the input from the SA layers. The key to the s-TCN is the way the SA layers and TCN residual unit are combined: the TCN-unit receptive field matches the size of the sliding block ($2S$) in the output of SA layers. 

As shown in Fig.~\ref{fig:system}(c) {\it right}, various 1D-FCN designs can then be used in the TCN unit. For instance, we set dilation rate to $1$ (effectively, no dilation since our block size is small), and use the kernel size $k_1=S/8$, with stride $s_1=S/8$. Note that in case of $k_1>s_1$, we pad by repeating the first frame; and for the last block, we repeat the last frame to fill in the block size. The convolved output is further passed through a stack-of-layers (SoL): weight-norm/ReLu/dropout, followed by 1D-FCN with $k_2, s_2= S/2, S/4$, effectively producing an embedding for every shift of $S$ frames. This is passed through SoL (see Sec.~\ref{sec:sa}), and a skip connection with a ReLu activation, keeping the output length the same as input. To obtain the FTM decisions, the sequence of embeddings ($B$) is propagated through the FCN+SoftMax layer, and the block-wise decision scores are averaged over time to produce the mitigation score ($Y$). Finally, the model parameters are learned by minimizing the XE loss in the discriminative branch, together with the CTC loss in the phonetic branch of the model.

\section{Experiments}
\label{sec:exps}
\subsection{Model Training and Inference}
The model trained only on the phonetic transcription task was used to initialize the the multi-task model in Fig.~\ref{fig:system}(a). The training data consist of 2700 hours of clean audio, further augmented with room impulse responses and echo residuals making a total of 8100 hours of data and transcribed with phonetic labels. As input, 40-D Mel-filterbank features at 100 fps are used, and the current frame is augmented with 3 neighbouring frames, resulting in 280-D features. For more details, see~\cite{Sigtia_2020, adya2020hybrid}. The network is comprised of 6 SA layers with 256 hidden units each, 4 heads within each block, and 1024 hidden units of a feed forward block. We use a block size B=64 frames (1.92s) with a shift of S=32 frames. In this way, the decisions are available $\sim$ every 1 sec. To train the FTM (discriminative) branch, we use $\sim$150k true and $\sim$30k negative triggers for each invocation type: VT ({\it set-I}), and TB ({\it set-II}). We further augment the training data by segmenting {\it set-I} VT data into the keyword and payload segments using the phone sequence predicted by VTD. The {\it payload} segments effectively enrich the {\it set-II} (TB-invocation) that do not contain the keyword. We used the Adam optimizer~\cite{kingma2014adam} with learning rate set to 5e-4, gradient clipping norm of 20, and 0.1 dropout rate (set via a cross-validation on 10\% of held-out data).

\subsection{Evaluation Procedure}
\label{sec:dps}
We focus on evaluating the accuracy and performance of the discriminative branch with different choices of streaming mechanisms for SSL (see Fig.~\ref{fig:system}). Specifically, we compare the newly introduced s-TCN mechanism with the streaming LSTM (s-LSTM) model, typically used for sequence classification tasks~\cite{hochreiter1997long}, and recently proposed for the VTD task~\cite{garg2021streaming}. We also compare with a vanilla Average layer (s-AVE), which consists of a fully-connected linear layer applied to each input frame, followed by ReLU. If not otherwise stated, we set the TCN kernel width/stride to 4/4 in the first convolutional filter (thus, non-overlapping frames), followed by the second kernel with width/stride 16/8, respectively. This ensures the receptive field/step size of the TCN aligns with the output of the SA layers. In s-LSTM approach, we set the size of LSTM hidden layers to 256, and use the average over the last 10 frames to get a decision score, as in~\cite{garg2021streaming}. We benchmark these models against the A2A FTM model, which uses LSTM-summary layer and non-streaming attention layers applied to the full input audio, to assess the difference in accuracy due to the streaming approximation. The target models are evaluated on in-house test sets for each invocation type, where the VT set contains 7.3k utterances (ratio 7:1 ratio of device-directed vs. undirected, mean duration 5.2$\pm$2.1 sec) and the TB set contains 38k utterances (ratio 3:1, duration 4.1$\pm$3.7 sec). Note that the target data is imbalanced, as it would be in the wild. 

\subsection{Results}
\label{sec:dps}

\subsubsection{Comparisons of Sequence Summary Layers}
\label{subsec:ss}
\begin{figure*}[t]
\begin{minipage}[b]{.25\linewidth}
  \centering
  \centerline{\includegraphics[width=5cm]{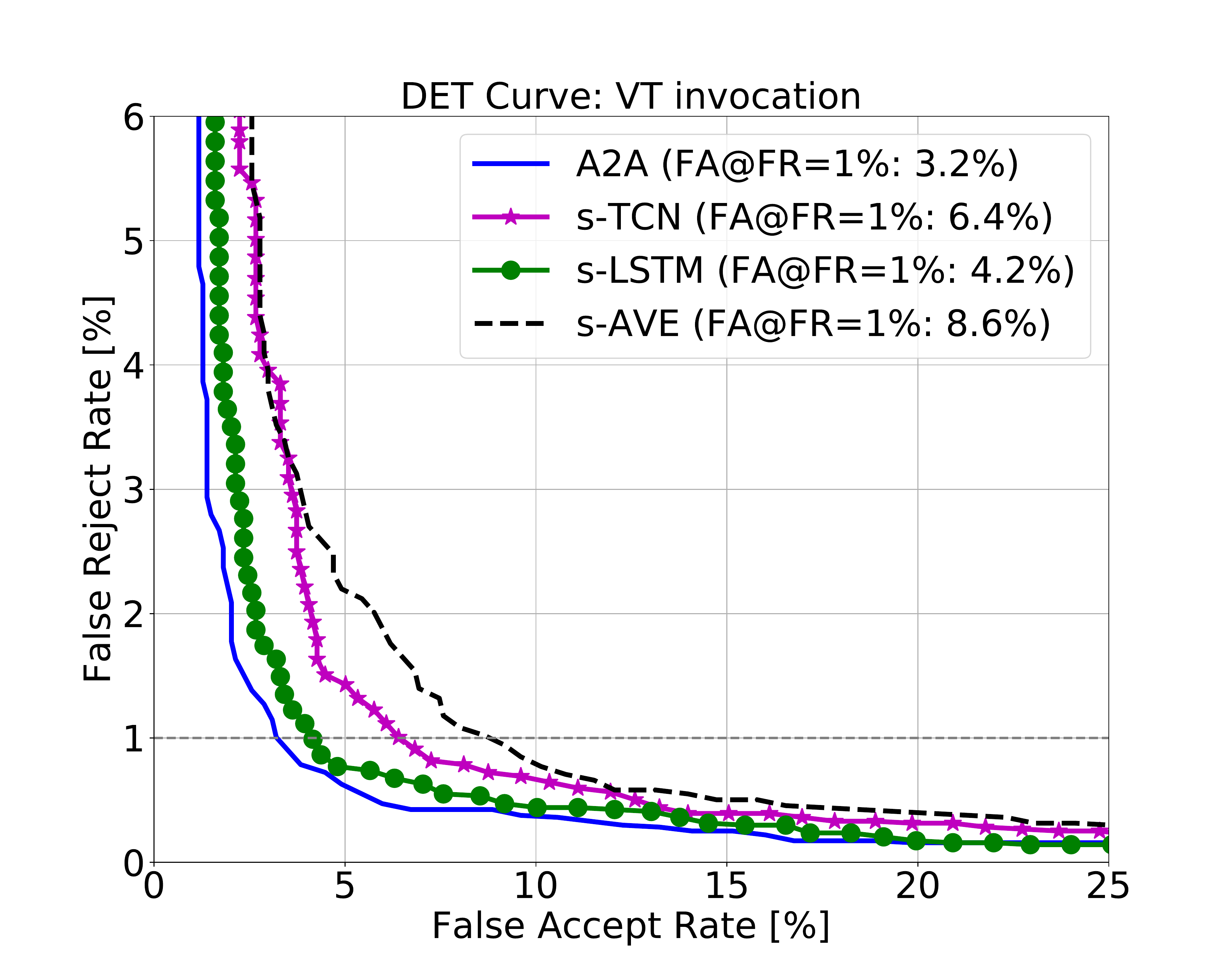}}
  \centerline{(a)}
\end{minipage}
\hfill
\begin{minipage}[b]{0.25\linewidth}
  \centering
  \centerline{\includegraphics[width=5cm]{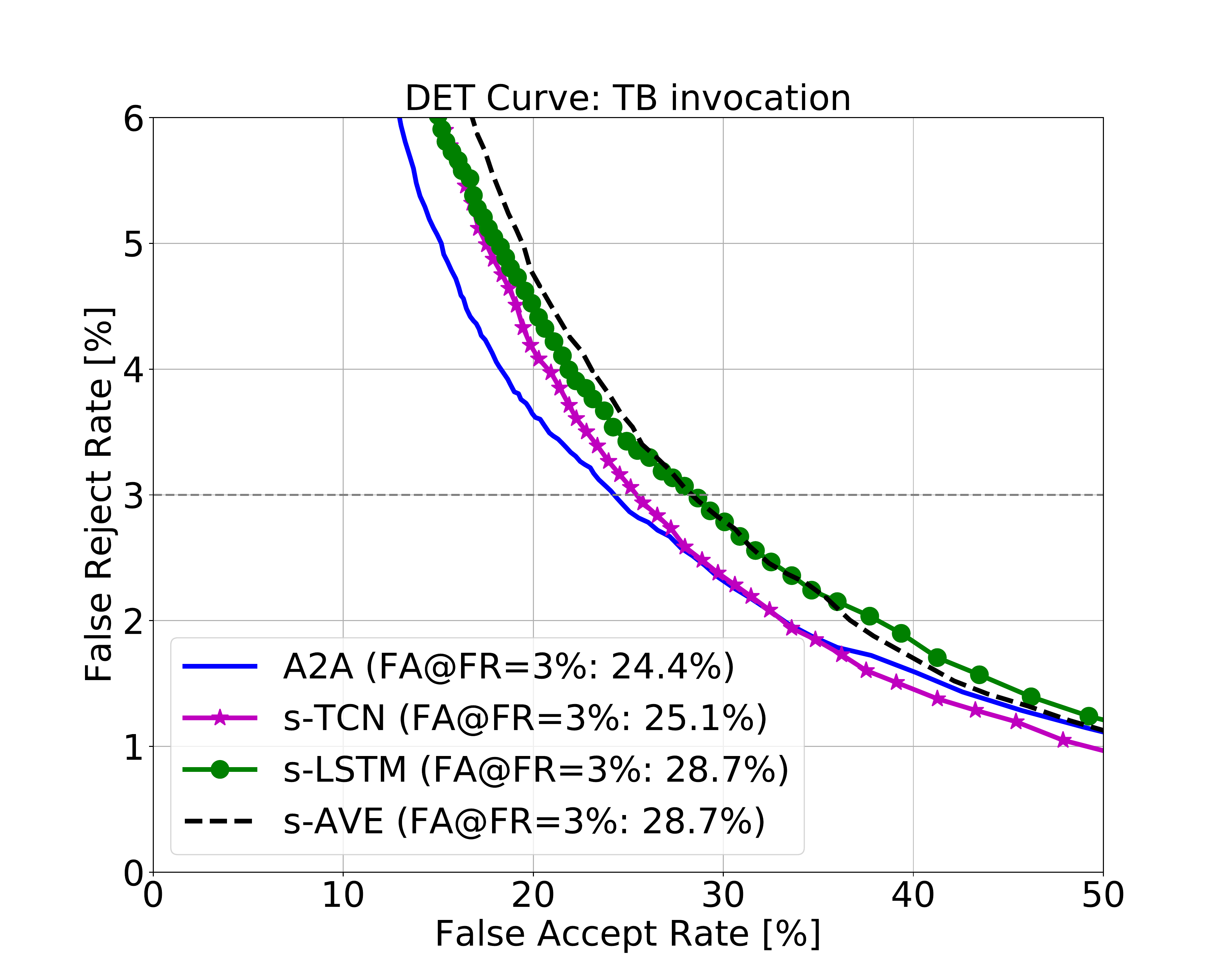}}
  \centerline{(b)}
\end{minipage}
\hfill
\begin{minipage}[b]{0.22\linewidth}
  \centering
  \centerline{\includegraphics[width=4.80cm]{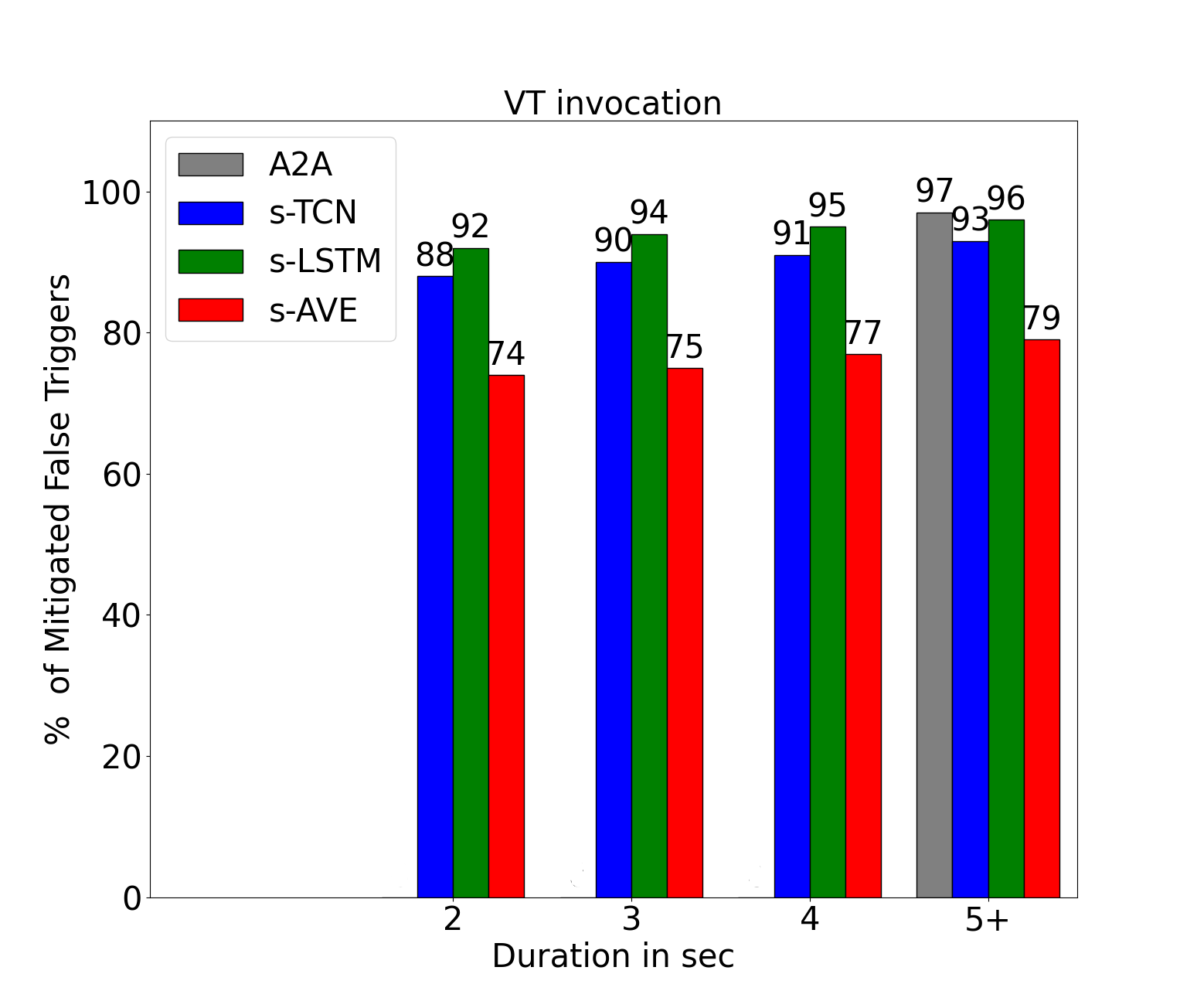}}
  \centerline{(c)}
\end{minipage}
\hfill
\begin{minipage}[b]{0.22\linewidth}
  \centering
  \centerline{\includegraphics[width=4.80cm]{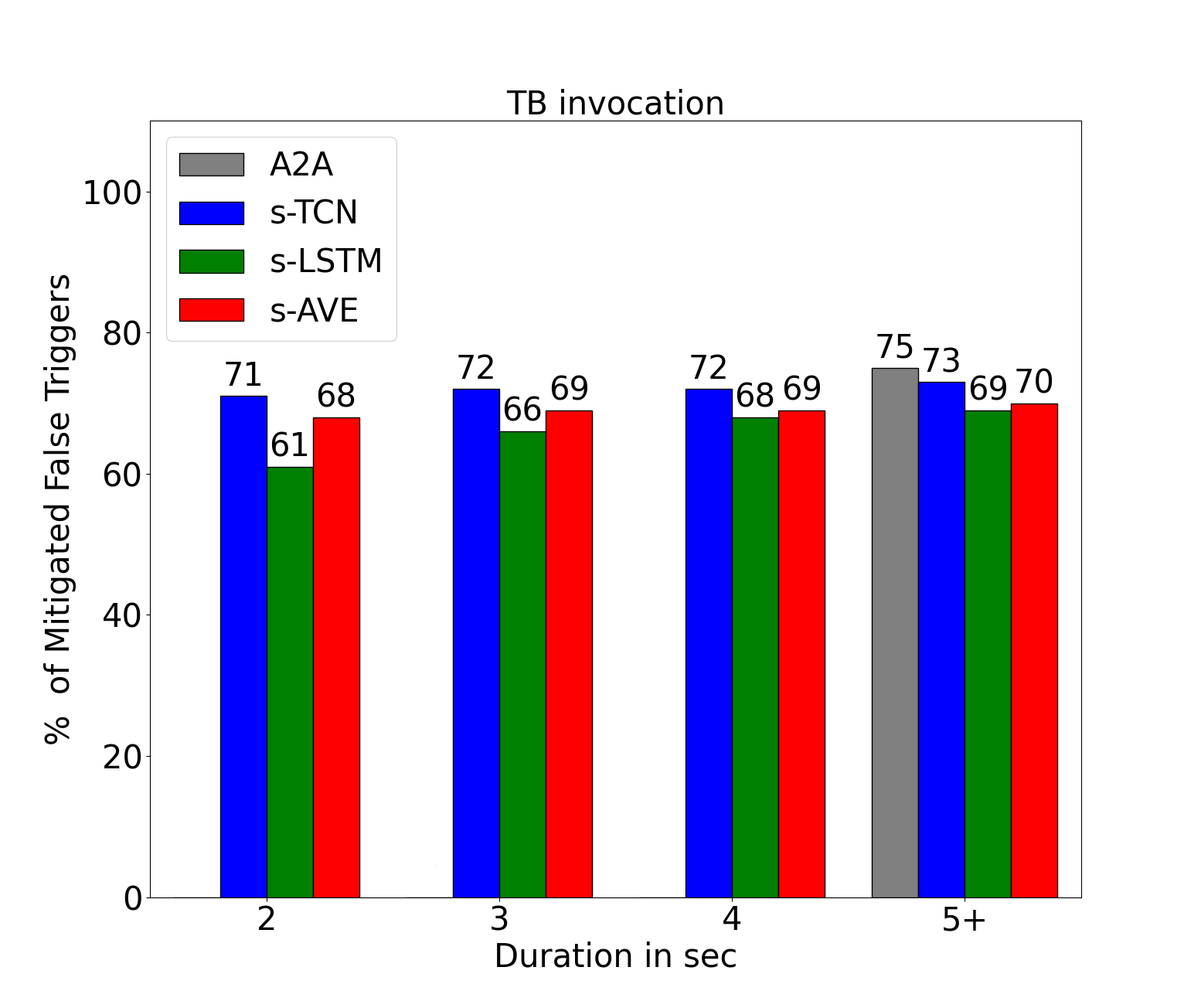}}
  \centerline{(d)}
\end{minipage}
\caption{Comparison of different streaming architectures.(a)-(b) DET curves depicting the selected operating region for each invocation. (c)-(d) The portion of correctly mitigated false triggers based on decision scores in time by the streaming models.}
\label{fig:DET}
\end{figure*}

Fig.~\ref{fig:DET} shows the results for the multi-invocation FTM systems that use a single model for both invocation types, and trained on the combined data sets {\it I+II}. Specifically, we report detection error trade-off (DET) curves (Fig.~\ref{fig:DET} (a)-(b)), by considering a hypothetical operating region for false reject rate (FRR) $<$ 5\%. The difference in false accepts (FAs) on the two invocation types is of an order of magnitude, highlighting the difficulty of FTM on the TB eval set. We report false accept rate (FAR) for VT at FRR=1\%, as in previous works~\cite{garg2021streaming, Sigtia_2020}, and for TB at FRR=3\%, thus being more flexible in terms of ralse rejects. First, we note that for the VT data, the s-LSTM model regresses slightly (from 3.2 to 4.2\% FAs) when compared to the A2A model, which has access to the full audio context. On the other hand, s-TCN seems to be less robust on this invocation type, yet, it consistently outperforms s-AVE. 


For the more challenging case of TB eval data (Fig.~\ref{fig:DET} (b)), we note that both s-LSTM and s-AVE produce similar and lower accuracy than the proposed s-TCN (25.1\% FAs), which almost matches that of the full A2A model. Even though FTM systems for VTD are known for being prone to the keyword hallucination (due to other similarly sounding words, which increases their FAs), the FTM accuracy on TB invocation has scarcely been investigated in literature. The fact that in this invocation type we get $\sim$ 5$\times$ more FAs than on the VT invocation is not surprising, and it highlights the difficulty of detecting device-directed speech for queries with no keyword. 

In Fig.~\ref{fig:DET} (c)-(d), we show the effect of early mitigation, i.e., when decisions are made after each shift of the streaming block ($\sim$ every 1 sec after the first 2 sec block). Specifically, we depict the proportion of correctly mitigated false triggers accumulated over time by each approach, and using the mitigation threshold for the operating points mentioned above. Note that in the VT case, both s-LSTM and s-TCN improve over time as more context is available, while s-AVE does not exhibit the same trend since it does not encode time (beyond the SA layers). Interestingly, in case of TB invocation, s-TCN outperforms the alternatives by large margin in terms of absolute accuracy and detection time. On the other hand, s-LSTM performs similarly to s-AVE on this task since there is no temporal event such as a keyword to focus on.



Table~\ref{tab:DATA} shows the accuracy and equal error rate (EER) of target FTM models trained on data from each invocation type separately (VT/TB) and when the multi-invocation (single) model is trained using both types of data (VT+TB). Note that EERs fall outside the operating region, making it difficult to compare target models; yet, we report it for reference. All models show inferior accuracy when trained on data of one invocation type and tested on another. Yet, we see that s-TCN is more robust in this setting. For instance, when train set {\it I} (VT augmented data) is used, s-TCN shows superior accuracy on TB eval data compared to the s-AVE layer (FAR=55.3\%). 

\begin{table}[]
  \centering
  \caption{The impact of training data on model accuracy. For VT and TB eval sets, we report FAR[\%] at FRR=1\% and 3\%, respectively. For reference, EER is reported in brackets.}
  \resizebox{0.65\textwidth}{!}{\begin{minipage}{\textwidth}
        \begin{tabular}{|l|l|l|l|l|l|l|}
        \hline
         {\it Test set}& \multicolumn{3}{c|}{\bf Voice-trigger (VT)} & \multicolumn{3}{c|}{\bf Touch-based (TB)} \\ \hline
        {\it Train set}&VT &TB &{\it VT+TB} &VT &TB &{\it VT+TB}      \\ \hline\hline
         {\bf \it A2A}   &   3.0 (1.9) &26.7 (7.2)   &3.2 (2.0)  &39.7 (11.4) &23.3 (10.5)  &24.3 (9.6)\\ \hline \hline
         {\bf s-TCN} &       6.6 (2.9) &22.2 (10.7) &6.4 (3.1)  &42.0 (12.9) &23.1 (10.1)  &{\bf 25.1} (9.7)       \\ \hline
         {\bf s-LSTM}&       4.3 (2.6) &47.8 (10.4) &{\bf 4.2} (2.6)  &43.3 (18.2) &28.7 (8.9)   &28.7 (9.6)       \\ \hline
         {\bf s-AVE} &       8.8 (3.4) &24.5 (10.4)  &8.6 (3.4)  &55.3 (20.5) &27.0 (10.4)  &28.7 (10.1)       \\ \hline
        \end{tabular}
      \end{minipage}}
    \label{tab:DATA}
\end{table}



\subsubsection{On-device Implications: Hardware Efficiency}
We next analyze the on-device performance in terms of (i) the peak memory consumption (PMC), and (ii) runtime latency (RTL), i.e., how long it takes for the model to produce the final decision given the audio input. In Table~\ref{tab:perf}, we quantify gains due to the streaming mechanisms in terms of relative reduction (\%) in PMC and RTL as compared with the baseline A2A model (no streaming). Note also that the computational complexity of the A2A model is $O(T^2D)$, where $T$ is the audio length, and $D$ is the embeddings size ($D=256$). Due to the streaming approximation, this becomes linear in time $O(TDS)$, where $S$ is the constant block shift. For comparisons, we chose a 4s long (device-directed, TB invocation) utterance recorded on a smart phone. From Table~\ref{tab:perf}, we note that s-TCN has largest gains over the A2A model (+$33.3\%$), while s-LSTM improves by +$7.3\%$ only. This is expected because s-TCN leverages its parallel computations (in contrast to a recursive LSTM), and 1D-convolutions (instead of the full $D{\times}D$ multiplications as in the s-AVE approach). The reduction in the PMC is particularly important for on-device deployment, especially on the low-power edge hardware such as wearables and smart speakers. In terms of RTL, we do not observe large differences between the streaming models, yet they all largely improve over the A2A model architecture (${40+}\%$). This is in support of streaming mechanisms despite the slight regression in models' accuracy. We also attribute the small differences in RTL to the fact that the size of the streaming block is fixed and relatively small (64 frames)\footnote{Thus, not enabling TCN to fully demonstrate its computational efficiency, as previously shown on longer sequences~\cite{bai2018empirical}.}. 

\begin{table}[]
\centering
\caption{Relative improvements in the model's peak memory consumption and runtime latency on a 2019 smart phone w.r.t. the A2A model evaluated on a 4s long utterance.}
\resizebox{0.9\textwidth}{!}{\begin{minipage}{\textwidth}
\begin{tabular}{|c|c|c|c|c|}
\hline
{\it Metric} & s-TCN & s-LSTM & s-AVE  \\ \hline
Peak memory consumption &  33.3\%&  7.3\%& 25.5\% \\ \hline
Runtime latency &  41.0\%&  41.5\%&  42.0\%\\ \hline
\end{tabular}
\end{minipage}}
\label{tab:perf}
\end{table}

\label{sec:ondev}

\section{Conclusions}
\label{sec:majhead}
We proposed a novel streaming on-device approach for detection of device-directed speech from both voice and touch-based invocations, and have provided an empirical analysis of accuracy and on-device performance for three streaming mechanisms. By combining data from both invocation types, we showed the feasibility of using a single model for both invocation types, despite a small regression in accuracy compared with invocation-specific models. This, in turn, enables us to have a single on-device model for both invocation types. Furthermore, while s-LSTM and s-TCN achieve similar accuracy on the voice-trigger invocation, and much higher than the vanilla Average layer, s-TCN is the most effective on the touch-based invocation. We also showed that s-TCN achieves large improvements over the s-LSTM in terms of the peak memory consumption on device and the mitigation decision time, which are critical for on-device deployment.

\bibliographystyle{IEEEbib}
{\small \bibliography{refs}}

\begin{thebibliography}{10}

\bibitem{bai2018empirical}
Shaojie Bai, J~Zico Kolter, and Vladlen Koltun,
\newblock ``An empirical evaluation of generic convolutional and recurrent
  networks for sequence modeling,''
\newblock {\em arXiv preprint arXiv:1803.01271}, 2018.

\bibitem{mallidi2018devicedirected}
Sri~Harish Mallidi, Roland Maas, Kyle Goehner, Ariya Rastrow, Spyros Matsoukas,
  and Björn Hoffmeister,
\newblock ``{Device-Directed Utterance Detection},''
\newblock in {\em Proc. Interspeech 2018}.

\bibitem{sainath2015convolutional}
Tara~N Sainath and Carolina Parada,
\newblock ``{Convolutional Neural Networks for Small-Footprint Keyword
  Spotting},''
\newblock in {\em Proc. Interspeech 2015}.

\bibitem{wu2018monophone}
Minhua Wu, Sankaran Panchapagesan, Ming Sun, Jiacheng Gu, Ryan Thomas, Shiv
  Naga~Prasad Vitaladevuni, Bjorn Hoffmeister, and Arindam Mandal,
\newblock ``{Monophone-Based Background Modeling for Two-Stage On-Device Wake
  Word Detection},''
\newblock in {\em Proc. ICASPP 2018}.

\bibitem{sigtia2020progressive}
Siddharth Sigtia, J.~Bridle, H.~Richards, P.~Clark, E.~Marchi, and Vineet Garg,
\newblock ``{Progressive Voice Trigger Detection: Accuracy vs Latency},''
\newblock {\em ArXiv}, vol. abs/2010.15446, 2020.

\bibitem{guo2018time}
Jinxi Guo, Kenichi Kumatani, Ming Sun, Minhua Wu, Anirudh Raju, Nikko
  Str{\"o}m, and Arindam Mandal,
\newblock ``Time-delayed bottleneck highway networks using a dft feature for
  keyword spotting,''
\newblock in {\em Proc. ICASSP 2018}.

\bibitem{Sigtia_2020}
Siddharth Sigtia, Pascal Clark, Rob Haynes, Hywel Richards, and John Bridle,
\newblock ``{Multi-Task Learning for Voice Trigger Detection},''
\newblock in {\em Proc. ICASSP 2020}.

\bibitem{cao21b_interspeech}
Songjun Cao, Yueteng Kang, Yanzhe Fu, Xiaoshuo Xu, Sining Sun, Yike Zhang, and
  Long Ma,
\newblock ``{Improving Streaming Transformer Based ASR Under a Framework of
  Self-Supervised Learning},''
\newblock in {\em Proc. Interspeech 2021}.

\bibitem{tsunoo2019transformer}
E.~{Tsunoo}, Y.~{Kashiwagi}, T.~{Kumakura}, and S.~{Watanabe},
\newblock ``{Transformer ASR with Contextual Block Processing},''
\newblock in {\em Proc. ASRU 2019}.

\bibitem{oh21_interspeech}
Yoo~Rhee Oh and Kiyoung Park,
\newblock ``{On-Device Streaming Transformer-Based End-to-End Speech
  Recognition},''
\newblock in {\em Proc. Interspeech 2021}.

\bibitem{wu2021transformer}
Chunyang Wu, Zhiping Xiu, Yangyang Shi, Ozlem Kalinli, Christian Fuegen, Thilo
  Koehler, and Qing He,
\newblock ``Transformer-based acoustic modeling for streaming speech
  synthesis,''
\newblock {\em Proc. Interspeech 2021}.

\bibitem{ellinas2020high}
Nikolaos Ellinas, Georgios~[..] Vamvoukakis, and Pirros Tsiakoulis,
\newblock ``High quality streaming speech synthesis with low,
  sentence-length-independent latency.,''
\newblock in {\em Proc. Interspeech 2020}.

\bibitem{tcn4kws}
Seungwoo Choi, Seokjun Seo, Beomjun Shin, Hyeongmin Byun, Martin Kersner,
  Beomsu Kim, Dongyoung Kim, and Sungjoo Ha,
\newblock ``Temporal convolution for real-time keyword spotting on mobile
  devices,''
\newblock in {\em Proc. Interspeech 2019}.

\bibitem{wang2021wake}
Yiming Wang, Hang Lv, D.~Povey, Lei Xie, and Sanjeev Khudanpur,
\newblock ``{Wake Word Detection with Streaming Transformers},''
\newblock {\em ArXiv}, vol. abs/2102.04488, 2021.

\bibitem{garg2021streaming}
Vineet Garg, Wonil Chang, Siddharth Sigtia, Saurabh Adya, Pramod Simha, Pranay
  Dighe, and Chandra Dhir,
\newblock ``Streaming transformer for hardware efficient voice trigger
  detection and false trigger mitigation,''
\newblock {\em Proc. Interspeech 2021}.

\bibitem{dighe2020knowledge}
Pranay Dighe, Erik Marchi, Srikanth Vishnubhotla, Sachin Kajarekar, and Devang
  Naik,
\newblock ``{Knowledge Transfer for Efficient On-device False Trigger
  Mitigation},''
\newblock in {\em Proc. ICASSP 2020}.

\bibitem{Tong2020StreamingRW}
Xiaosu Tong, Che-Wei Huang, Sri~Harish Mallidi, Shaun Joseph, Sonal Pareek,
  Chander Chandak, A.~Rastrow, and R.~Maas,
\newblock ``{Streaming ResLSTM with Causal Mean Aggregation for Device-Directed
  Utterance Detection},''
\newblock {\em ArXiv}, vol. abs/2007.09245, 2020.

\bibitem{vaswani2017attention}
Ashish Vaswani, Noam Shazeer, Niki Parmar, Jakob Uszkoreit, Llion Jones,
  Aidan~N Gomez, \L~ukasz Kaiser, and Illia Polosukhin,
\newblock ``{Attention is All You Need},''
\newblock in {\em Proc. NeurIPS 2017}.

\bibitem{dai2019transformerxl}
Zihang Dai, Z.~Yang, Yiming Yang, J.~Carbonell, Quoc~V. Le, and
  R.~Salakhutdinov,
\newblock ``{Transformer-XL: Attentive Language Models Beyond a Fixed-Length
  Context},''
\newblock in {\em Proc. ACL 2019}.

\bibitem{NEURIPS2018_e44fea3b}
Hu~Liu, Sheng Jin, and Changshui Zhang,
\newblock ``{Connectionist Temporal Classification with Maximum Entropy
  Regularization},''
\newblock in {\em Proc. NeurIPS 2018}, 2018, vol.~31.

\bibitem{adya2020hybrid}
Saurabh Adya, Vineet Garg, Siddharth Sigtia, Pramod Simha, and Chandra Dhir,
\newblock ``{Hybrid Transformer/CTC Networks for Hardware Efficient Voice
  Triggering},''
\newblock in {\em Proc. Interspeech 2020}.

\bibitem{lea2016temporal}
Colin Lea, Rene Vidal, Austin Reiter, and Gregory~D Hager,
\newblock ``Temporal convolutional networks: A unified approach to action
  segmentation,''
\newblock in {\em Proc. ECCV 2016}.

\bibitem{kiranyaz20211d}
Serkan Kiranyaz, Onur Avci, Osama Abdeljaber, Turker Ince, Moncef Gabbouj, and
  Daniel~J Inman,
\newblock ``1d convolutional neural networks and applications: A survey,''
\newblock {\em Mechanical systems and signal processing}, vol. 151, pp. 107398,
  2021.

\bibitem{kingma2014adam}
Diederik~P Kingma and Jimmy Ba,
\newblock ``Adam: A method for stochastic optimization,''
\newblock {\em arXiv preprint arXiv:1412.6980}, 2014.

\bibitem{hochreiter1997long}
Sepp Hochreiter and J{\"u}rgen Schmidhuber,
\newblock ``{Long short-term memory},''
\newblock {\em Neural computation}, vol. 9, no. 8, pp. 1735--1780, 1997.

\end{thebibliography}

\end{document}